# SMILE: a universal tool for modulated-enhanced localization microscopy to achieve minimal three-dimensional resolution

Hongfei Zhu[2,6], Yile Sun[1,6], Xinxun Yang[1], Enxing He[1], Lu Yin[5], Hanmeng Wu[1], Mingxuan Cai[1], Yubing Han[1], Renjie Zhou[2,*], Cuifang Kuang[1,3,4,*], Xu Liu[1,3,4]

[1]*State Key Laboratory of Extreme Photonics and Instrumentation, College of Optical Science and Engineering, Zhejiang University, Hangzhou 310027, China*

[2]*Department of Biomedical Engineering, The Chinese University of Hong Kong, Hong Kong, China*

[3]*ZJU-Hangzhou Global Scientific and Technological Innovation Center, Hangzhou, Zhejiang 311200, China*

[4]*Collaborative Innovation Center of Extreme Optics, Shanxi University, Taiyuan, Shanxi 030006, China*

[5]*College of Optical and Electronic Technology, China Jiliang University, Hangzhou, Zhejiang 310018, China*

[6]*These authors contributed equally to this work*

[*]*Correspondence and requests for materials should be addressed to C.K. (e-mail: cfkuang@zju.edu.cn) and R.Z. (e-mail: rjzhou@cuhk.edu.hk)*

## Abstract

Modulation-enhanced localization microscopy (MELM) has demonstrated significant improvements in both lateral and axial localization precision compared to conventional single-molecule localization microscopy (SMLM). However, lateral modulated illumination based MELM (MELM*xy*) remains fundamentally limited to two-dimensional imaging. Here we present three-dimensional Single-Molecule Modulated Illumination Localization Estimator (SMILE) that synergistically integrates lateral illumination modulation with point spread function engineering. By simultaneously exploiting lateral modulation patterns and an accurate point spread function (PSF) model for 3D localization, SMILE achieves near-theoretical-minimum localization uncertainty, demonstrating an average 4-fold enhancement in lateral precision compared to conventional 3D-SMLM. Crucially, SMILE exhibits exceptional compatibility with diverse PSFs and different illumination patterns with various structures including 4Pi configurations, making it a versatile tool that can be easily adapted for different experimental setups. When integrated with 4Pi microscopy, 4Pi-SMILE shows particular promise for achieving sub-10 nm axial resolution and approaching isotropic resolution. From the simulations and proof-of-concept experiments, we verified the superiority of SMILE over 3D-SMLM and ordinary MELM. We highlight SMILE as a novel methodology and robust framework that holds great potential to significantly promote the development of MELM.

## Introduction

Over the past two decades, super-resolution fluorescence microscopy has revolutionized the way of imaging bio-macromolecules – including proteins, lipids, filaments – by overcoming the diffraction barrier that previously obscured subcellular structural details. These technological breakthroughs are inextricably linked to fundamental discoveries in biomedical research. Among super-resolution modalities, single-molecule localization microscopy (SMLM)[1-3] relies on detecting the emission of a single fluorophore within a sub-diffraction region by randomly switching on/off the fluorescent molecule , then fitting the centroids of all emission spots. Many efforts have been made to extend SMLM to three dimensional (3D) applications. The most commonly used approach for 3D-SMLM is based on engineering the point spread functions (PSFs), such as introducing astigmatism by inserting a single cylindrical lens into the imaging path[4]. Other PSFs, such as a double helix[5], phase ramp[6], or tetrapod[7], can be obtained using phase masks, deformable mirrors, or other optical devices. Complementary solutions, such as two focal planes (Biplane configuration[8]) are also used to obtain axial information by analyzing the intensity variations in different focal planes of the same molecule. Nevertheless, the axial localization precision of PSF engineering-based SMLM is usually 2-3 times worse than its lateral precision. To address this critical limitation, interferometric detection schemes have been proposed, such as interferometric photoactivated localization microscopy (iPALM)[9,10] and 4Pi single marker switching nanoscopy (4Pi-SMS)[11,12], which consist of 4Pi configuration (two opposing objectives) and allow fluorescent signals to interfere and divide into three or four channels. Compared to single-objective SMLM, the 4Pi configuration provides an approximate sixfold improvement in axial localization precision, due to the axial fluorescence interference pattern, with slightly improvement in lateral precision from the increased photon collection by the two objective lenses.

In recent years, the advent of Minimal fluorescence flux (MINFLUX)[13] has broken through the paradigm of SMLM, achieving Ångström-level localization precision with a lower photon budget by utilizing a doughnut-shaped PSF. Building upon this conceptual framework, modulation-enhanced localization microscopy (MELM) with laterally modulated illumination (referred to as MELM*xy*), including ROSE[14], SIMPLE[15], and SIMFLUX[16], leverages structured illumination patterns to achieve approximately a twofold lateral precision enhancement over standard 2D-SMLM. MELM with axial modulation (referred to as MELM*z*), exemplified by ROSEz[17] and ModLoc[18], providing approximately a six-fold improvement in axial localization precision. In these methods, the spatial location of molecules in each dimension is determined by estimating their relative spatial displacement within a phase-shifting fringe pattern. Notably, the MELM approaches reported recently have primarily concentrated on optimizing either lateral resolution or axial resolution, without achieving simultaneous enhancements of both, resulting in suboptimal spatial isotropy, and the MELM*xy* methods are only reported as 2D modality.

Here we propose single molecule modulated illumination localization estimator (SMILE) as a solution to achieve 3D imaging of MELM, which considers both lateral modulation information and 3D point spread function model for 3D localization. Standard 3D-SMLM with PSF engineering suffers from rapid deteriorations in lateral localization precision when dealing with defocusing emitters. SMILE significantly improves the lateral localization precision at out-of-focus positions, with an average enhancement of more than 3.7 times over 3D-SMLM with an astigmatic PSF. To achieve this, we developed an optimal joint fitter that is accelerated by a graphics processing unit (GPU), reaching the theoretical Cramér-Rao lower bound (CRLB). Notably, the SMILE optimal fitter can be extended to various illumination schemes and PSF geometries, especially when applying 4Pi configuration, 4Pi-SMILE achieves unparalleled isotropic three-dimensional resolution down to sub-10 nm. We believe SMILE would unlock new possibilities for quantitative 3D mapping of subcellular structures with ultra-high resolution and open up a new avenue for biomedical research.

## Results

### Principle

The basic concept of SMILE is shown in Fig. 1a. The sample is excited by axially uniform 3D interferometric sinusoidal patterns during its on-state. Subsequent localization in each direction is achieved by employing three phase shifts of 120°; Thus, a total of six excitation patterns were sufficient for lateral localization (three phase-shifted patterns per orthogonal axis). The axial localization is determined by PSF geometries (PSF engineering or interferometric detection schemes).

Considering that the diffusion of PSF at defocused positions is much more significant than that at the focus plane, leading to a degradation of lateral localization precision. Luckily, the lateral modulation scheme of SMILE will have a more significant improvement effect when the feature size is larger. Hence, Ast SMILE (Astigmatism SMILE, SMILE with PSF engineering) can compensate for the loss of localization precision caused by the defocused PSF, making the precision at defocusing positions comparable to that in the focus plane.

Compared to recent MELM*xy* methods, SMILE offers four significant improvements. First, whereas recent MELM*xy* techniques employ a two-dimensional (2D) PSF with Gaussian profile during the fitting process and obtain only 2D coordinates without axial information, SMILE utilizes pre-calibrated 3D cubic-spline interpolated PSF models for fitting. This enables 3D imaging without introducing systematic error that may arise from adopting a Gaussian approximation. Second, SMILE substantially enhances the lateral localization precision at defocused positions, exceeding a two-fold improvement, and overcomes the limited lateral resolution inherent to PSF engineering. Third, to attain optimal localization precision, SMILE employs a loss function based on maximum likelihood estimation (MLE), implements Levenberg–Marquardt (L-M) algorithm to minimize the loss function, and approaches Cramér-Rao lower bound (CRLB) regardless of the excitation or detection scheme used. Additionally, SMILE leverages the computational power of graphics processing units (GPUs) to enable parallel computation and accelerate the analysis process, making the method suitable for a wide range of applications. These features make SMILE highly versatile, enabling researchers to obtain accurate and reliable results across diverse experimental conditions.

The flow chart of SMILE and corresponding detailed descriptions can be found in Supplementary Fig. 1. The most crucial aspects include the cubic-spline representation of the experimental 3D PSF and the optimal joint fitting algorithm, which are elaborated in detail in **Methods**.

## Single molecule simulations

Single molecule simulations were conducted to verify the feasibility of SMILE. Single molecule images were simulated through two imaging modalities: astigmatic PSF model (Ast-SMILE) and 4Pi-PSF model (4Pi-SMILE). For Ast-SMILE, each localization was assigned 5000 photons (distributed across 6 sub-images) with 5 background photons/pixel (totaling 30 photons/pixel). For 4Pi-SMILE, the same photon count per localization was maintained, but with 2 background photons/pixel (totaling 48 photons/pixel). Astigmatic data is fitted with both SMILE and state-of-art 3D-SMLM algorithm[19] (SMAP2018). When fitting with SMLM algorithm, all six sub-images are summed up as one single image. While the 4Pi data processing pipeline of SMILE involves a multi-step fitting approach: initially, the data are integrated across the four detection channels and subjected to SMILE fitting to determine the lateral coordinates, followed by integration across the six excitation patterns and global fitting to extract the axial position. Subsequently, all 24 sub-images are collectively integrated and analyzed using the SMAP2018 algorithm to generate the 3D-SMLM localization results. In this simulation, single-molecule images were generated from the cubic-spline representation of the PSF and multiplied by different numbers of photons for each sub-image according to the relative position of the single molecule with respect to the sinusoidal patterns or different 4Pi channels. Each sub-image was then added with a constant background and then degraded by Poisson noise. The modulation depth in both $x$, $y$ direction is set as 0.95 and the period of the illumination pattern is 220 nm. 2D

PSFs from several axial positions of the astigmatic PSF model and the 4Pi-PSF are path, the fluorescence signal can be either collect by single objective with a deformable mirror for pupil engineering (Ast SMILE) or two objective lenses with 4Pi detection modulation module (4Pi-SMILE), then imaged by an EMCCD or sCMOS camera, resulting in six corresponding sub-images within a single exposure for Ast SMILE or twenty four (six patterns times four 4Pi channels) sub-images within a single exposure for 4Pi-SMILE. **b** Basic principles of 3D- and 4Pi- SMILE. In both 3D- and 4Pi-SMILE, the fluorophore is modulated by excitation of two groups of orthogonal sinusoidal patterns (three in *x* direction and three in *y* direction). The phase difference between adjacent fringes with the same direction is 120 degrees. While in 4Pi-SMILE, the emission pattern of the fluorophore is further modulated by the 4Pi interference cavity, resulting in the generation of four distinct intensity channels for each sub-image. The specific intensity distribution across these channels serves as a sensitive indicator of the fluorophore's axial position within the sample. **c1** Normalized PSFs (4Pi and astigmatism) with different depths used in single molecule simulations, scale bar: 500 nm. **c2, c3** *x*, *y, z* localization improvement of 4Pi- (first panel) and Ast SMILE (second panel) over 3D-SMLM, both experimental and theoretical (numerical CRLB).

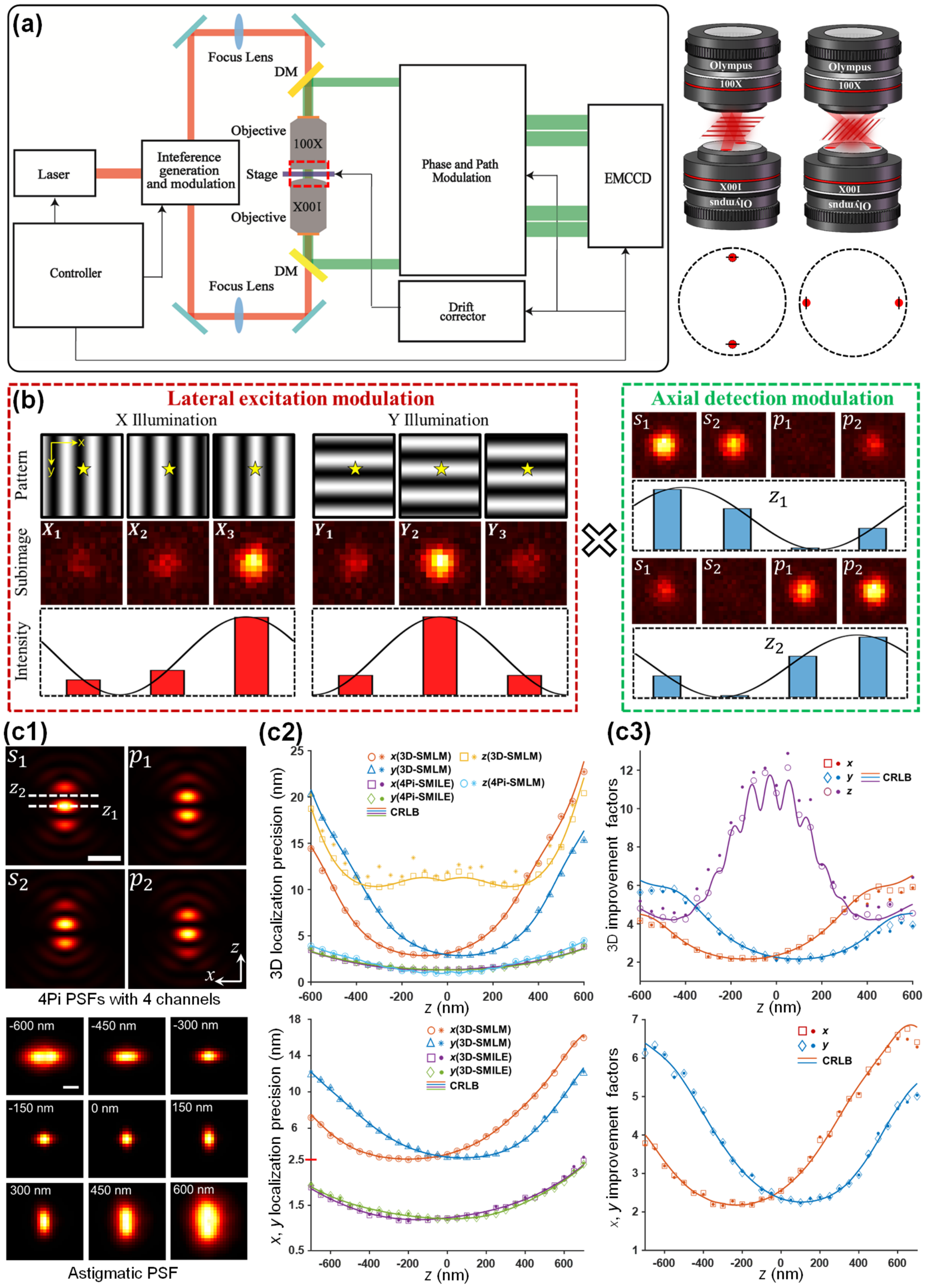

**Fig. 1** Principle and simplified setup of SMILE. **a** Simplified setup of SMILE. The excitation laser was modulated and interfered at the sample plane from a single objective lens (either upper or lower) to create an axially uniform illumination pattern. But various illumination patterns can be achieved through the modulation and interference module, excited from single or dual objective lens. SMILE can reconstruct under different illumination schemes and here we only illustrate SMILE with lateral illumination modulation. The phase of the interference fringes is modulated to equally shift 3 times along *x*/*y* directions. In the imaging

shown in Fig. 1 c1 for reference. We then evaluated the *x*, *y*, *z* localization precisions at different axial positions between fitted 3D positions (SMILE and 3D-SMLM, respectively) and ground truth of the 2000 simulated molecules at each axial position (symbols). Furthermore, we calculated the root-mean-square error between the fitted coordinates and ground truth (asterisks/dots). The localization precision as well as the localization accuracy of the fitted positions of both SMILE and 3D-SMLM achieves the estimated Cramér–Rao lower bound (CRLB) (denoted as dashed lines for 3D-SMLM and solid lines for SMILE) under astigmatic PSF model and 4Pi-PSF model (Fig. 1 c2). The theoretical lateral precision/accuracy improvement factors (solid lines) and simulation experiment improvement factors (symbols for precision improvement and asterisks/dots for accuracy improvement) of SMILE over 3D-SMLM using astigmatic PSF model and 4Pi-PSF model are shown in Fig. 1 c3), respectively. The results indicate that the lateral precision/accuracy improvement of SMILE is in good agreement with the theoretical values when fitting with astigmatic PSF. The average improvement of localization precision/accuracy in *x* and *y* direction is > 3.7 folds for both astigmatic PSF and 4Pi-PSF, and > 8 folds in axial direction for 4Pi-PSF with the help of 4Pi detection.

The theoretical performance of SMILE under different signal-to-noise ratio,

modulation depth as well as period of the modulation pattern can be found in Supplementary Fig. 2~4.

**3D super-resolution reconstruction on simulated nuclear pore complex.**

We simulated a nuclear pore complex (NPC) raw data (Fig. 2) with axially homogeneous modulated illumination (i.e., the excitation patterns at different depths will remain consistent. Supplementary Fig. 6a is the schematic diagram of 3D sinusoidal pattern in *x* and *y* directions respectively) using an experimental astigmatic PSF model (Fig. 1(c1)) and comparing the results between Astigmatism SMILE (Ast-SMILE), 4Pi-SMILE and 3D-SMLM. The total number of photons of each fluorescent molecule is set to be 3000 with 120 background photons/pixel in total. The modulation depth of both *x* and *y* patterns is set as 0.95 and the period of the interference pattern is 219.3 nm (Numerical aperture: 1.45, excitation wavelength: 560 nm). The fluorophores are distributed from -600 nm to 600 nm in the simulations. In the simulations of the nuclear pore complex, the polygons were assigned a radius of 110 nm. It was observed that in regions with large out-of-focus depths, the structure of the polygons was difficult to resolve using 3D-SMLM (Fig. 2b), whereas Ast SMILE was able to resolve the structure easily (Fig. 2a). The multi-modal Gaussian fitting on the NPC profile (Fig. 2 a1) showed that the average distance between adjacent loci is 47.9 nm, which indicates that the circumference is 383.2 nm and the diameter is approximately 122 nm. To further verify the improvement of Ast-SMILE in lateral resolution, we also conducted simulations of free 3D curves (as shown in Supplementary Fig. 5), with the same simulation parameters as mentioned above (i.e., total photon number, background noise, pitch, and modulation depth of the excitation pattern). It is also important to note that under axially homogeneous modulated illumination, the axial resolution of Ast-SMILE does not exceed that of 3D-SMLM, however, thanks to the interference mechanism of 4Pi configuration, 4Pi-SMILE exhibits outstanding axial super-resolution performance (Fig. 3e and f). Even under low signal-to-noise ratio (SBR) conditions, 4Pi-SMILE can still distinguish the axial bilayer structure of Nup96 sites, and its axial performance is still excellent even at the defocused planes. As shown in the Fig.2 g – i, the axial distances between the three NPC are 47.7 nm, 50.3 nm and 49.2 nm, which are consistent with the theoretical value.

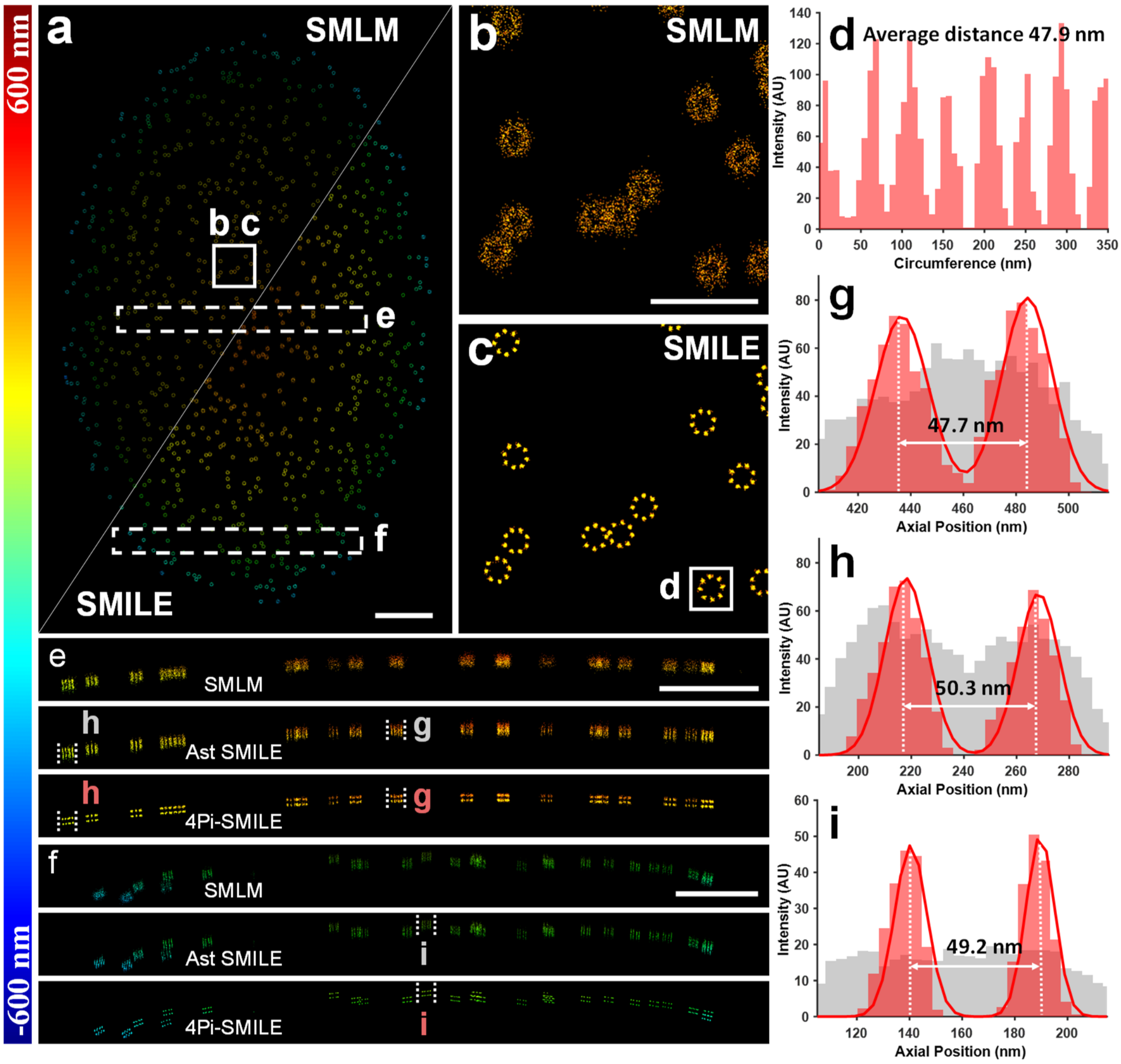

**Fig. 2** NPC simulation results. **a** Color-coded (depth) 3D images resulting from SMILE and 3D-SMLM, respectively. **b, c** SMLM and SMILE results of the white solid box area in **a**. **d** Circular frofile of the NPC of the white solid box in **c**. **e, f** Side-view cross-sections of white dashed area shown in **aa**. **g – i** Axial profiles of white dotted area shown in **e** and **f**. Scale bar: 2 μm for **a**, 200 nm for **b**, 1 μm for**e** and **f**.

**3D super-resolution reconstruction on simulated microtubules.**

Microtubule data were also simulated to verify the performance of SMILE over 3D-SMLM. The ground truth of the microtubules was obtained from the EPFL 2016 SMLM Challenge training data set (http://bigwww.epfl.ch/smlm). Other simulation parameters, such as total photon number, background noise, and modulation depth and pitch, were consistent with those in NPC simulations. The hollow structure of the microtubules was easily distinguishable in both Ast-SMILE and 4Pi-SMILE (Fig. 3 a and c), but not in 3D-SMLM (Fig. 3 b), indicating an improvement in lateral resolution. By multi-modal Gaussian fitting of the lateral profile of the microtubules (Fig. 3 d), the fitted diameters of the hollow structures were 14.8 nm and 14.4 nm for inner diameters and 29.54 nm and 37.77 nm for outer diameters, close to the ground truth (~22 nm for average diameter). 4Pi-SMILE also exhibited a significant improvement in axial resolution over 3D-SMLM and Ast-SMILE when 4Pi configuration was applied. As shown in Fig. 3 g – j, the axial hollow structure of microtubules can be discerned with an inner diameter of 15.7 nm and an outer diameter of 40.6 nm where both 3D-SMLM and Ast-SMILE cannot distinguish. It can be proved that 4Pi-SMILE with interferometric detection scheme demonstrated simultaneous improvements in axial and lateral resolution.

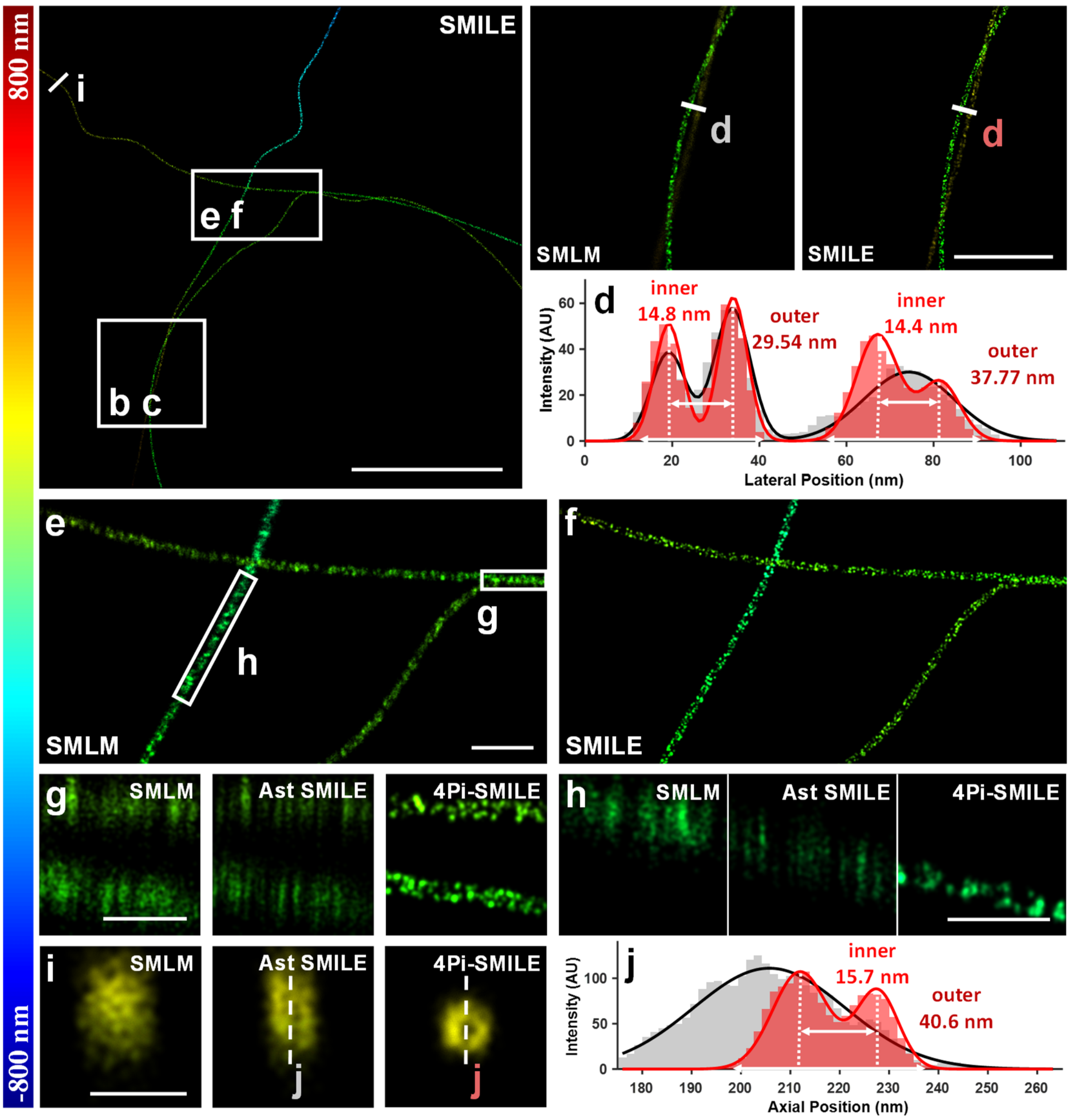

**Fig. 3** Microtubules simulations. **a** Color-coded (depth) 3D images resulting from SMILE

and 3D-SMLM, respectively. **b, c** SMLM and SMILE results of the white solid box area in **a**. **d** Lateral profiles of white dotted area shown in **b** and **c**. **e, f** SMLM and SMILE results of the white solid box area in **a**. **g, h** Side-view cross-sections along the lines shown in **e**. **i** Side-view cross-sections along the lines shown in **a**. **j** Axial profiles of white dotted area shown in **i**. Scale bars: 2 μm for **a**, 500 nm for **c**, 200 nm for **e**, 100 nm for **g** and **h**, 50 nm for **i**.

**Experimental results.**

To evaluate the performance of Ast SMILE, experiments on 40 nm fluorescent nanospheres were conducted as proof of concept. We added an astigmatism aberration to the imaging system, and calibrated the PSF in advance (Fig. 4a) is the PSF at different depths used in this experiment. 40 nm fluorescent beads were settled on the upper surface of the cover slip (Fig. 4c) and staged to different depths (-450nm to 450nm in 100nm steps in the z direction) to capture raw data with modulated illumination in *x* directions (50 repeated sets for each depth). The period is around 317.8 nm and the average modulation contrast is 0.77 across the field of view (FOV). Image sequences with 3 different modulation phases are used to calculate the localization precision $\Delta x$ of Ast SMILE (standard deviation of *x* position of the same molecule) and the sequences are summed up to calculate the localization precision $\Delta x$ of 3D-SMLM. No significant drift was found during the collection of experimental data. We selected four candidate points that are basically at the same depth from the FOV (Fig. 4b) and calculated localization precision curves for each depth. Through the scatter-line plot, Ast SMILE shows an impressive improvement on both *x* over 3D-SMLM (Fig. 4 d-g). The achieved average localization precision over the whole imaging depth for the 3 candidate molecules are 5.8nm, 5.5nm, 5.3 nm and 6.1 nm for Ast SMILE in contrast to 12.9 nm, 11.5 nm, 10.9 nm and 12.0 nm for 3D-SMLM, indicating an improvement around 2.1 folds in average and nearly 3 folds for positions with large defocus depth. The improvement can be much higher if the modulation depth can reach larger than 0.9 and the modulation period can be smaller.

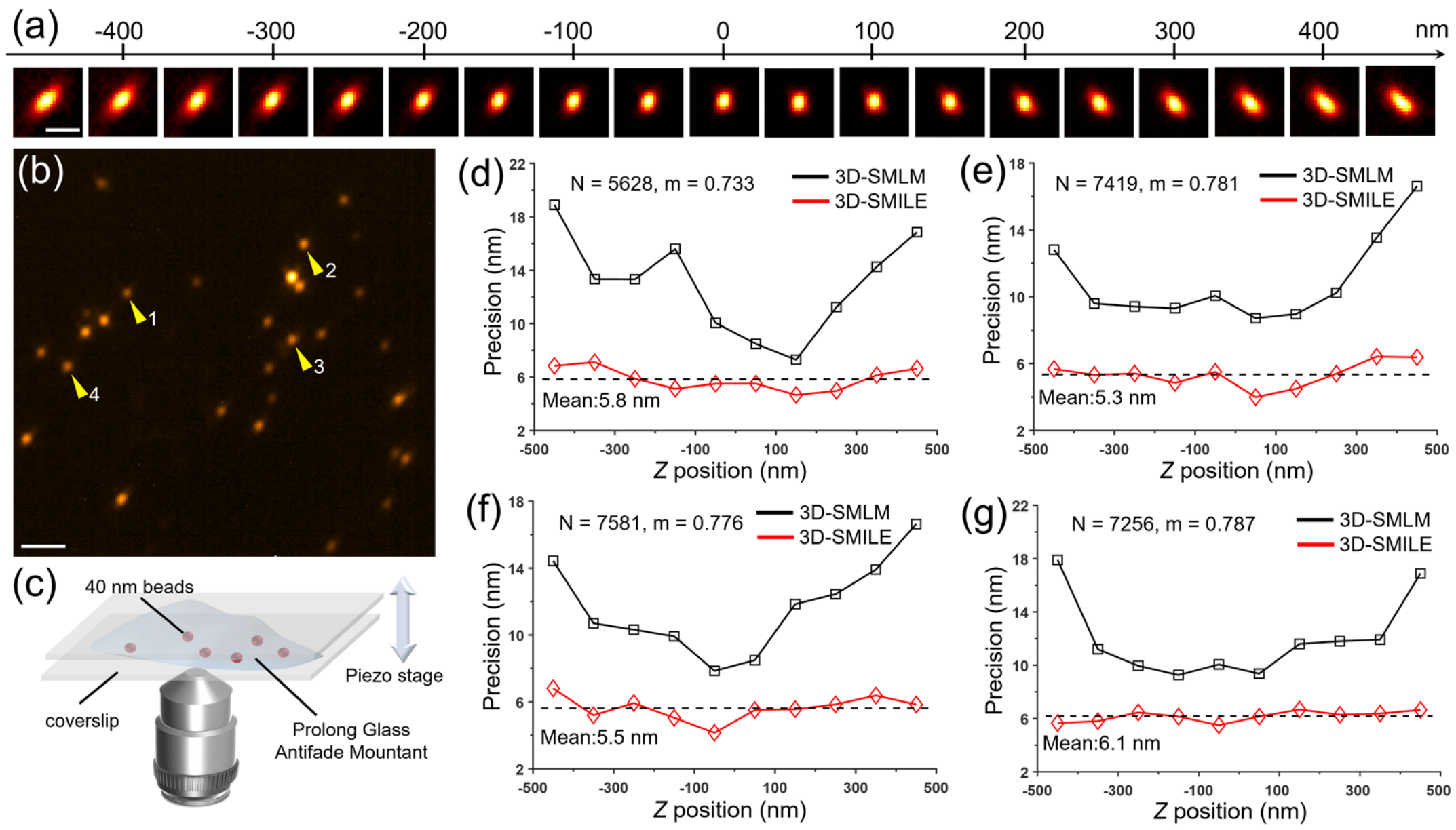

**Fig. 4** Experimental validation of Ast SMILE. **a** Pre-calibrated normalized experimental PSFs with different depths. **b** Partial FOV showing the sample with sparse single beads. The localization precision of the single molecules pointed by the yellow arrows are measured. **c** A sketch of the sample used to measure localization precision at different depths. The 40 nm particles are on the upper surface of the cover slip and moved by a Piezo stage to different axial depths. **d~g** Localization precision curve across the depth of field of the four candidate molecules (pointed by yellow arrows and numbered 1 to 4). The mean photon numbers are 5628, 7419, 7581 and 7256 with background noise approximately109.1 photons/pixels. The mean modulation depth is 0.733, 0.781, 0.778 and 0.787, respectively. Scale bars: 2 um for **b** and 500 nm for others.

**Discussion.**

It has been demonstrated that SMILE can significantly enhance the lateral localization precision compared to 3D-SMLM when introducing sinusoidal interference illumination. In addition to this improvement, when applying 4Pi configuration, SMILE can also be used for interferometric detection scheme, resulting in near-isotropic 3D resolution enhancement. It is expected that modulated illumination will continue to lead to further improvements in the resolution of 3D single-molecule techniques, offering a powerful tool for more advanced biomedical discoveries.

## Acknowledgments

This work was financially sponsored by the National Natural Science Foundation of China (62125504, 62361166631); STI 2030—Major Projects (2021ZD0200401); the Fundamental Research Funds for the Central Universities (2022FZZX01-20).

## Competing financial interests

The authors declare no competing interests.

## References

1 Betzig, E. *et al.* Imaging intracellular fluorescent proteins at nanometer resolution. *science* **313**, 1642-1645 (2006).

2 Hess, S. T., Girirajan, T. P. & Mason, M. D. Ultra-high resolution imaging by fluorescence photoactivation localization microscopy. *Biophys J* **91**, 4258-4272, doi:10.1529/biophysj.106.091116 (2006).

3 Rust, M. J., Bates, M. & Zhuang, X. Sub-diffraction-limit imaging by stochastic optical reconstruction microscopy (STORM). *Nature methods* **3**, 793-796 (2006).

4 Huang, B., Wang, W., Bates, M. & Zhuang, X. Three-dimensional super-resolution imaging by stochastic optical reconstruction microscopy. *Science* **319**, 810-813, doi:10.1126/science.1153529 (2008).

5 Pavani, S. R. *et al.* Three-dimensional, single-molecule fluorescence imaging beyond the diffraction limit by using a double-helix point spread function. *Proc Natl Acad Sci U S A* **106**, 2995-2999, doi:10.1073/pnas.0900245106 (2009).

6 Baddeley, D., Cannell, M. B. & Soeller, C. Three-dimensional sub-100 nm super-resolution imaging of biological samples using a phase ramp in the objective pupil. *Nano Research* **4**, 589-598, doi:10.1007/s12274-011-0115-z (2011).

7 Shechtman, Y., Sahl, S. J., Backer, A. S. & Moerner, W. E. Optimal point spread function design for 3D imaging. *Phys Rev Lett* **113**, 133902, doi:10.1103/PhysRevLett.113.133902 (2014).

8 Juette, M. F. *et al.* Three-dimensional sub–100 nm resolution fluorescence microscopy of thick samples. *Nature Methods* **5**, 527-529, doi:10.1038/nmeth.1211 (2008).

9 Shtengel, G. *et al.* Interferometric fluorescent super-resolution microscopy resolves 3D cellular ultrastructure. *Proc Natl Acad Sci U S A* **106**, 3125-3130, doi:10.1073/pnas.0813131106 (2009).

10 Wang, G., Hauver, J., Thomas, Z., Darst, S. A. & Pertsinidis, A. Single-Molecule Real-Time 3D Imaging of the Transcription Cycle by Modulation Interferometry. *Cell* **167**, 1839-1852 e1821, doi:10.1016/j.cell.2016.11.032 (2016).

11 Aquino, D. *et al.* Two-color nanoscopy of three-dimensional volumes by 4Pi detection of stochastically switched fluorophores. *Nat Methods* **8**, 353-359, doi:10.1038/nmeth.1583 (2011).

12 Huang, F. *et al.* Ultra-High Resolution 3D Imaging of Whole Cells. *Cell* **166**, 1028-1040, doi:10.1016/j.cell.2016.06.016 (2016).

13 Balzarotti, F. *et al.* Nanometer resolution imaging and tracking of fluorescent molecules with minimal photon fluxes. *Science* **355**, 606-612, doi:10.1126/science.aak9913 (2017).

14 Gu, L. *et al.* Molecular resolution imaging by repetitive optical selective exposure. *Nature*

Methods **16**, 1114-1118, doi:10.1038/s41592-019-0544-2 (2019).

15 Reymond, L. *et al.* SIMPLE: Structured illumination based point localization estimator with enhanced precision. *Opt Express* **27**, 24578-24590, doi:10.1364/OE.27.024578 (2019).

16 Cnossen, J. *et al.* Localization microscopy at doubled precision with patterned illumination. *Nat Methods* **17**, 59-63, doi:10.1038/s41592-019-0657-7 (2020).

17 Gu, L. *et al.* Molecular-scale axial localization by repetitive optical selective exposure. *Nat Methods* **18**, 369-373, doi:10.1038/s41592-021-01099-2 (2021).

18 Jouchet, P. *et al.* Nanometric axial localization of single fluorescent molecules with modulated excitation. *Nature Photonics* **15**, 297-304, doi:10.1038/s41566-020-00749-9 (2021).

19 Li, Y. *et al.* Real-time 3D single-molecule localization using experimental point spread functions. *Nature Methods* **15**, 367-369, doi:10.1038/nmeth.4661 (2018).

# Methods

**Calculation of cubic-spline interpolated PSFs.** The 3D PSF can be divided into many stacked voxels and each voxel $(i,j,k)$ can be described by a 3D cubic spline as a summation of multiple ternary polynomials as follows[20],

$$f_{i,j,k}(x,y,z) = \sum_{l=0}^{3}\sum_{m=0}^{3}\sum_{n=0}^{3} a_{i,j,k,l,m,n}\left(\frac{x-x_l}{\Delta x}\right)^l\left(\frac{y-y_m}{\Delta y}\right)^m\left(\frac{z-z_n}{\Delta z}\right)^n \tag{1}$$

where $\Delta x$ and $\Delta y$ are the pixel sizes of the PSF in the object space in $x$ and $y$ directions, respectively; and $\Delta z$ is the step size in $z$ direction. $x_l$, $y_m$ and $z_n$ are the start positions of the voxel $(i,j,k)$. For each voxel, there are 64 coefficients $a_{i,j,k,l,m,n}$ to be determined, thus 64 linear equations are needed to solve the coefficients. To get a sufficient number of linear equations, one can firstly up-sample the original PSF stack by 3 times in all 3 directions, respectively. Then there will be 64 up-sampled coordinates inside and on the boundary of each voxel, which can be directly used to calculate the cubic spline coefficients.

Cubic spline representation can accurately describe various PSFs and another benefit of representing the PSF formulaically is to calculate the partial derivative of the PSF more straightforwardly. The expressions of partial derivatives of $f_{i,j,k}(x,y,z)$ with respect to $x,y,z$ are:

$$\frac{\partial f_{i,j,k}}{\partial x} = \sum_{l=1}^{3}\sum_{m=0}^{3}\sum_{n=0}^{3} a_{i,j,k,l,m,n}\frac{l}{\Delta x}\left(\frac{x-x_l}{\Delta x}\right)^{l-1}\left(\frac{y-y_m}{\Delta y}\right)^m\left(\frac{z-z_n}{\Delta z}\right)^n \tag{2}$$

$$\frac{\partial f_{i,j,k}}{\partial y} = \sum_{l=0}^{3}\sum_{m=1}^{3}\sum_{n=0}^{3} a_{i,j,k,l,m,n}\frac{m}{\Delta y}\left(\frac{x-x_l}{\Delta x}\right)^{l}\left(\frac{y-y_m}{\Delta y}\right)^{m-1}\left(\frac{z-z_n}{\Delta z}\right)^n \tag{3}$$

$$\frac{\partial f_{i,j,k}}{\partial z} = \sum_{l=0}^{3}\sum_{m=0}^{3}\sum_{n=1}^{3} a_{i,j,k,l,m,n}\frac{n}{\Delta z}\left(\frac{x-x_l}{\Delta x}\right)^{l}\left(\frac{y-y_m}{\Delta y}\right)^{m}\left(\frac{z-z_n}{\Delta z}\right)^{n-1} \tag{4}$$

This enables us to quickly calculate the relevant partial derivatives when applying some fitting algorithms based on gradient calculation.

**IAB-based model of 4Pi-PSF.** To localize the axial position with optimal precision which approaches Cramér–Rao lower bound (CRLB). We utilize IAB-based 4Pi-PSF model, the state-of-art method to fitting 4Pi-PSF. Compare with the original solution which extracts axial positions from photometry, IAB model considers the information from the fluorescence interference and change the multi-channel 3D PSF to a 4D PSF with x, y, z, and φ. The IAB model can be written as:

$$P(x,y,z,\varphi) = I + A\cos(\varphi) + B\cos(\varphi) \quad (5)$$

Where $I(x,y,z) = |E_1|^2 + |E_2|^2$, $A(x,y,z) = 2(Re(E_1)Re(E_2) + Im(E_1)Im(E_2))$, $B(x,y,z) = 2(Re(E_1)Im(E_2) - Im(E_1)Re(E_2))$. Here, $E_1$ and $E_2$ are amplitude field of fluorescent signals from the upper and lower objective lens, respectively. $\varphi$ is the interference phase which is depend on the optical path difference of fluorescence and the axial position of the fluorescent molecule. The specific 3D matrixes to describe $I$, $A$, and $B$ can be obtained through 3D scan of the two fo the 4Pi-PSFs from four channels with known phase retardancy. In SMILE experiments, the similar Levenberg–Marquardt algorithm can be used for MLE. With the help of IAB-based model, the axial localization precision can reach theoretical minimum. It should be noticed that, although IAB-based model possesses the potential to localized the molecules which beyond one fluorescence interference period in the axial direction, we still introduce vertical astigmatism of 60 mrad to implement phase unwrapping. Because under the condition of low photon count and low signal-to-background ratio, distinguishing the interference period by IAB-based model may still be inaccurate.

**Optimal joint fitting algorithm.** Here we firstly describe the forward imaging model of each sub-image. For 3 sub-images modulated by sinusoidal patterns in *x* direction, there intensities can be written as:

$$I_x^i = \frac{N_x}{3} \times \left(1 + m_x \sin\left(\overrightarrow{p_x} \cdot (x_0, y_0, z_0) + (i-1) \times \frac{2\pi}{3} + \varphi_x\right)\right) \times PSF(x - x_0, y - y_0, z_0) + bg_x^i \quad (6)$$

where $i = 1, 2, 3$ denotes sub-images with different phase shifts, $\overrightarrow{p_x}$ denotes direction and period vector of the fringes, $\varphi_x$ denotes the global phase of the sinusoidal pattern, $m_x$ is the modulation depth of the fringes, $N_x$ is the total photon number of 3 sub-images in *x* direction, $bg$ is the background noise, and $(x_0, y_0, z_0)$ is the coordinate of fluorescent molecule. The other 3 sub-images modulated by sinusoidal patterns in y direction are arranged in a similar way:

$$I_y^j = \frac{N_y}{3} \times \left(1 + m_y \sin\left(\overrightarrow{p_y} \cdot (x_0, y_0, z_0) + (j-1) \times \frac{2\pi}{3} + \varphi_y\right)\right) \times PSF(x - x_0, y - y_0, z_0) + bg_y^j \quad (7)$$

where $j = 1, 2, 3$ denotes sub-image with different phase shifts in *y* direction. In the fitting process, $\overrightarrow{p_x}$, $\overrightarrow{p_y}$, $\varphi_x$, $\varphi_y$ are regarded as known (pattern parameters are pre-estimated, see step 6 of Ast SMILE workflow, Supplementary Fig. 1). The parameters remaining to be determined are $[N_x, N_y, m_x, m_y, x_0, y_0, z_0, bg_x^{1\sim3}, bg_y^{1\sim3}]$, and a loss function based on MLE is constructed to solve all the parameters. MLE is the method of choice for fitting data degrades by Poisson noise and the loss function of MLE is given as follows[21]:

$$\chi_{MLE}^2 = 2\sum_{i=1}^{3}\left(\sum_k\left(\mu_k^i - x_k^i\right) - \sum_k x_k^i ln\left(\frac{\mu_k^i}{x_k^i}\right)\right) + 2\sum_{j=1}^{3}\left(\sum_k\left(\mu_k^j - x_k^j\right) - \sum_k x_k^j ln\left(\frac{\mu_k^j}{x_k^j}\right)\right) \quad (8)$$

where $\mu_k^i$ and $\mu_k^j$ are the expected number of photons in pixel $k$ of *x* sub-images and *y* sub-images which can be derived from forward imaging model combined with cubic-spline interpolated PSF model and $x_k^i$, $x_k^j$ are the measured number of photons.

To minimize $\chi_{MLE}^2$, Levenberg–Marquardt (L-M) algorithm is a choice which has widely used in least-squares fitting for its low time complexity and robustness. In L-M algorithm[19] for (7), the search direction of parameter $\theta_r$ is given by (see Supplementary note 1 for detailed expressions of partial derivatives with respect to each undetermined parameter):

$$\left(H_{r,s} + \lambda I\right)\Delta\theta_r = J_r \quad (9)$$

where $\lambda$ is the damping factor and $H_{r,s}$ is the Hessian matrix without the second partial derivatives term. $I$ is a diagonal matrix equal to the diagonal elements of the Hessian matrix which is defined as:

$$H_{r,s} = \sum_{i=1}^{3} \sum_k \frac{\partial \mu_k^i}{\partial \theta_r} \frac{\partial \mu_k^i}{\partial \theta_s} \frac{x_k^i}{{\mu_k^i}^2} + \sum_{j=1}^{3} \sum_k \frac{\partial \mu_k^j}{\partial \theta_r} \frac{\partial \mu_k^j}{\partial \theta_s} \frac{x_k^j}{{\mu_k^j}^2} \quad (10)$$

$J_r$ is the Jacobian matrix, defined as:

$$J_r = \sum_{i=1}^{3} \sum_k \frac{\partial \mu_k^i}{\partial \theta_r} \frac{(x_k^i - \mu_k^i)}{\mu_k^i} + \sum_{j=1}^{3} \sum_k \frac{\partial \mu_k^j}{\partial \theta_r} \frac{(x_k^j - \mu_k^j)}{\mu_k^j} \quad (11)$$

When $\lambda$ is large enough, the L-M algorithm will behave as gradient descent methods. To ensure the convergence of the L-M algorithm, in this work, $\lambda$ will be increased by a factor of 10 if an iteration step does not decrease $\chi^2_{MLE}$ or $H_{r,s}$ is not positive definite and will be decreased by a factor of 10 if the iteration step successfully decreases $\chi^2_{MLE}$.

**GPU acceleration.** The GPU implementation of the optimal joint fitter in Ast SMILE follows a framework similar to that described in literature. The data and global parameters are saved in the GPU global memory, and each thread is responsible for completing all calculations for one molecule candidate. Each block consists of 64 threads. The fitter achieves a fitting speed of >300,000 fits per second on a server with an Intel I9-13900k CPU, NVIDIA 4090 24G GPU, and 128G RAM, which is sufficient for online analysis and real-time imaging. Even on a common personal laptop with an i7-9750H CPU, GTX 1660ti GPU, and 16G RAM, the fitting speed exceeds 10,000 fits per second.

**Sample preparation.** For 40nm-bead samples, we used the dark red fluorescent FluoSpheres beads (F8789, excitation/emission maxima = 660/680 nm) as samples. Dilute the beads in ultrapure water at a ratio of 1:10000, and vibrate in an ultrasonic cleaner for 5 mins. Next, 200 μl of the diluted beads were placed on a coverslip and allowed to stand for 10 mins. The absorbent tissue was then used to remove excess water and Prolong GlassDiamond Antifade (Thermo Fisher Scientific, Inc.) before sealing the coverslip.

**Postprocessing.** Color-coded depth processing code is mainly written using MATLAB, drawing on previous literature[22]. The color bar from cold to warm reflects the relative axial position of each localization. As for volumetric display of reconstructed results, a commercial software named Vutara SRX Viewer from Bruker was used. By importing 3D coordinate of each localization for 3D-SMLM and Ast SMILE respectively, both 3D distributions of reconstructed results and the corresponding improvement in localization precision can be fully demonstrated and reflected.

# References

19 Li, Y. *et al.* Real-time 3D single-molecule localization using experimental point spread functions. *Nature Methods* **15**, 367-369, doi:10.1038/nmeth.4661 (2018).

20 Babcock, H. P. & Zhuang, X. Analyzing Single Molecule Localization Microscopy Data Using Cubic Splines. *Scientific Reports* **7**, 552, doi:10.1038/s41598-017-00622-w (2017).

21 Laurence, T. A. & Chromy, B. A. Efficient maximum likelihood estimator fitting of histograms. *Nat Methods* **7**, 338-339, doi:10.1038/nmeth0510-338 (2010).

22 Xu, F. *et al.* Three-dimensional nanoscopy of whole cells and tissues with in situ point

spread function retrieval. *Nat Methods* **17**, 531-540, doi:10.1038/s41592-020-0816-x (2020).

# Supplementary Information

## Supplementary Figure 1: Flow chart of SMILE

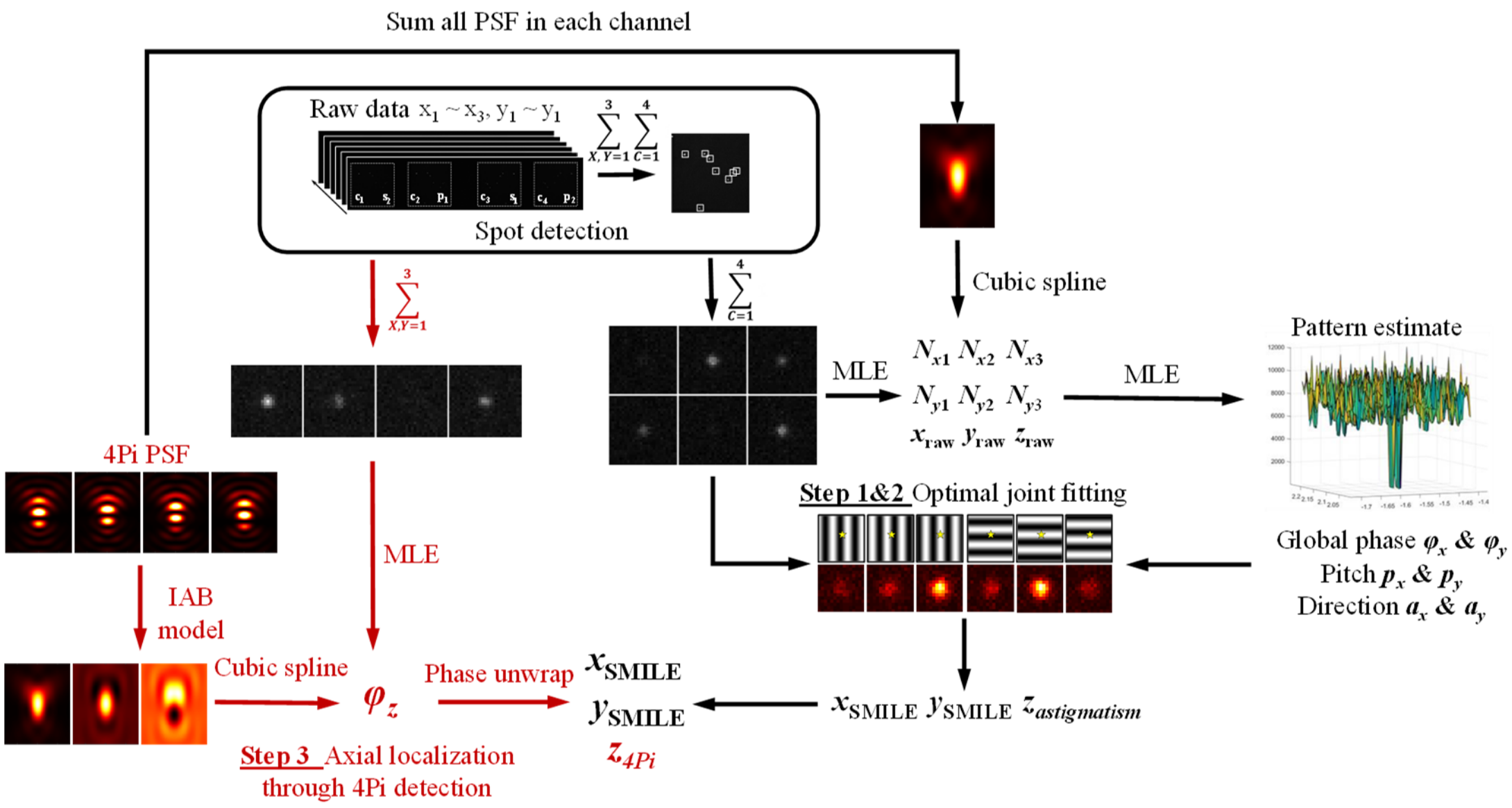

When processing Ast SMILE data, after calibrating the experimental PSF using discrete beads, the PSF image stack is then interpolated using cubic spline (see details in the Methods) and a differentiable three-dimensional PSF model can be generated. After ROI detection and sub-image partition, a maximum likelihood estimation (MLE) based fitter with ratiometric calculation of photon numbers in each of the six sub-images is applied (Supplementary Note 2). This fitter also provides rough position estimates, while the ratiometric calculation provides relative location with respect to interference fringes. The rough position estimates and relative locations are used to calculate the fringe parameters, i.e., direction, period and phase (Supplementary Note 2). Then, a GPU based optimal joint fitting algorithm will take into comprehensive consideration of the fringe parameters and the information of all 6 sub-images to further refine the initial rough position. Briefly speaking, this optimal fitter utilizes the MLE loss function and applies the Levenberg–Marquardt (LM) iterative scheme to minimize the loss function (see details in the Methods). Ultimately, SMILE image can be obtained after 3D mapping. When applying 4Pi-SMILE, the IAB model would be utilized to generate differentiable four-dimensional PSF model. And through cubic spline and MLE, the more precise axial localization can be obtained, which enables isotropic three-dimensional resolution. And the axial results of Ast SMILE can be further used to conduct phase unwrapping and prevent period aliasing caused by fluorescence interference.

**Supplementary Figure 2: Localization precision of 3D-SMLM and Ast SMILE under different signal-to-noise ratio**

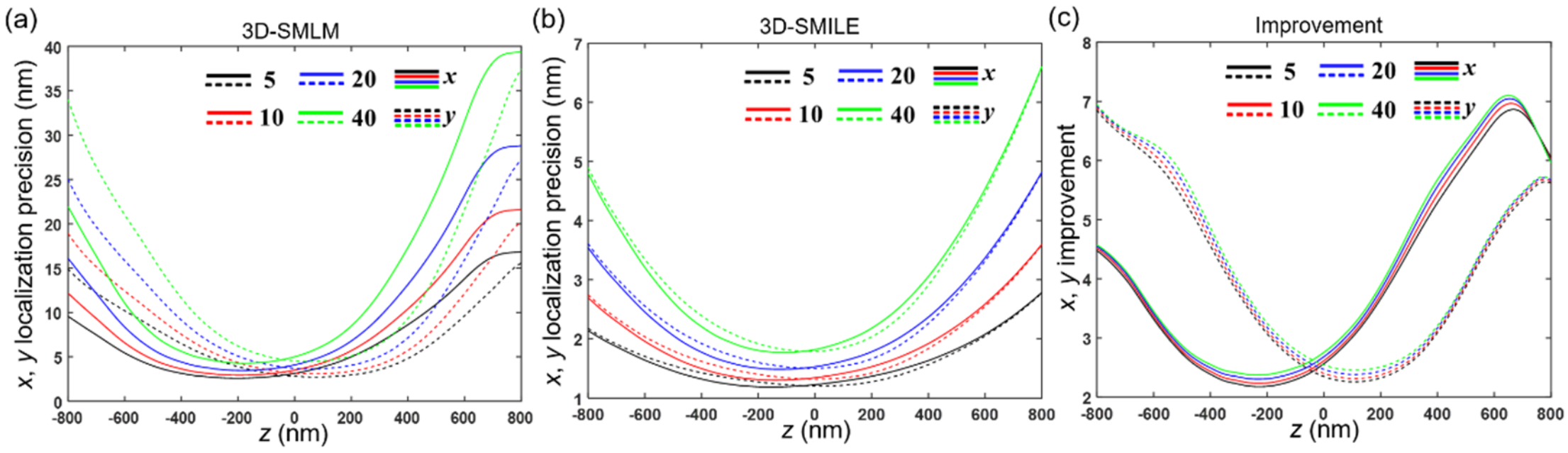

The theoretical localization precision of 3D-SMLM (panel **a**) and Ast SMILE (panel **b**) using an astigmatic PSF (Fig. 2a) has been calculated. The total photon number of each localization is 5000 and the background photon noise is varying from 5 photons/pixel to 40 photons/pixel. The modulation depth of the modulated sinusoidal pattern is 0.95 and the period of the pattern is set as 220 nm. While localization precision will deteriorate with the increase of background noise, the improvement of Ast SMILE over 3D-SMLM (panel **c**) changes very few with respect to the signal-to-noise ratio.

## Supplementary Figure 3: Localization precision of 3D-SMLM and Ast SMILE under different modulation depth

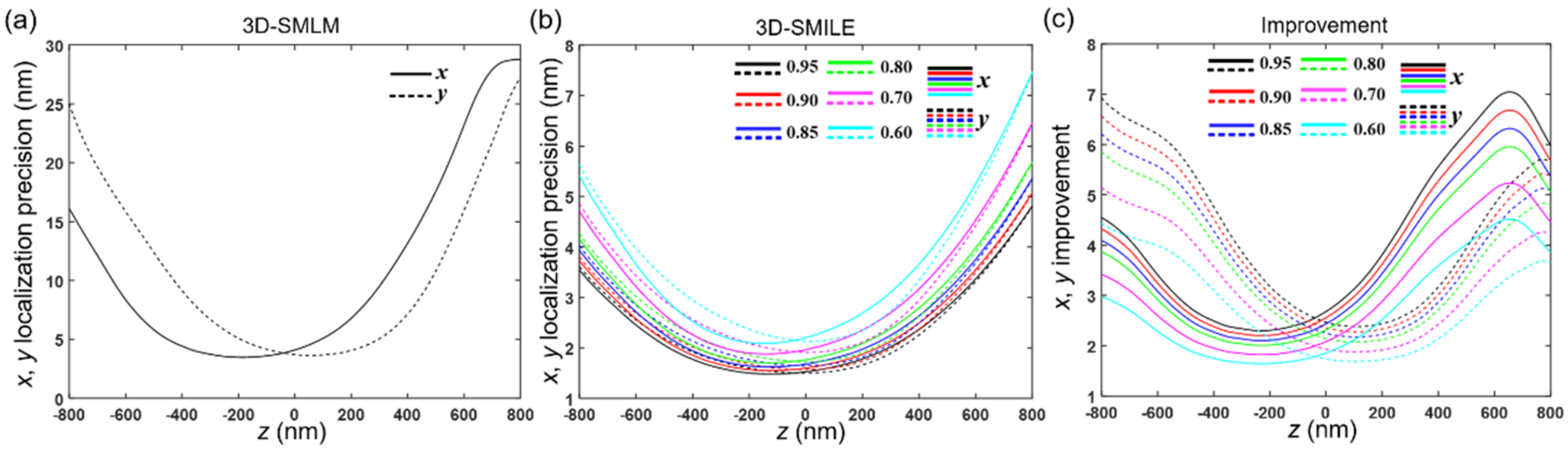

The theoretical localization precision of 3D-SMLM (panel **a**) and Ast SMILE (panel **b**) using an astigmatic PSF (Fig. 2a) has been calculated. The localization precision of 3D-SMLM is uncorrelated with the modulation depth. The total photon number of each localization is 5000 and the background photon noise is 20 photons/pixel. The modulation depth of the modulated sinusoidal pattern is varying from 0.6 to 0.95 and the period of the pattern is set as 220 nm. The results indicate the modulation depth is better larger than 0.8 to ensure a good performance of Ast SMILE.

.

**Supplementary Figure 4: Localization precision of 3D-SMLM and Ast SMILE under different modulation period**

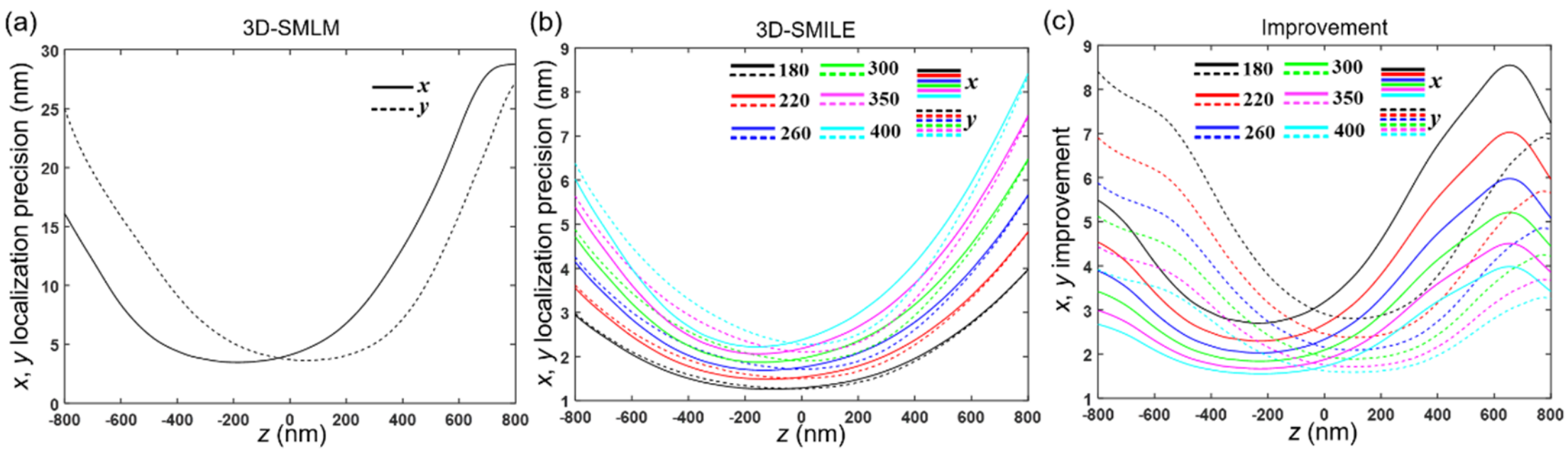

Localization precision of 3D-SMLM and Ast SMILE under different modulation period. The theoretical localization precision of 3D-SMLM (panel **a**) and Ast SMILE (panel **b**) using an astigmatic PSF (Fig. 2a) has been calculated. The localization precision of 3D-SMLM is uncorrelated with the modulation period. The total photon number of each localization is 5000 and the background photon noise is 20 photons/pixel. The modulation depth of the modulated sinusoidal pattern is set as 0.95 and the period of the pattern is varying from 180 nm to 400 nm. The results indicate a nonlinear relation between the pitch of the pattern and improvement. To ensure an optimal performance, the interference beam is preferably grazing incidence.

## Supplementary Figure 5: Reconstruction results of 3D free curve

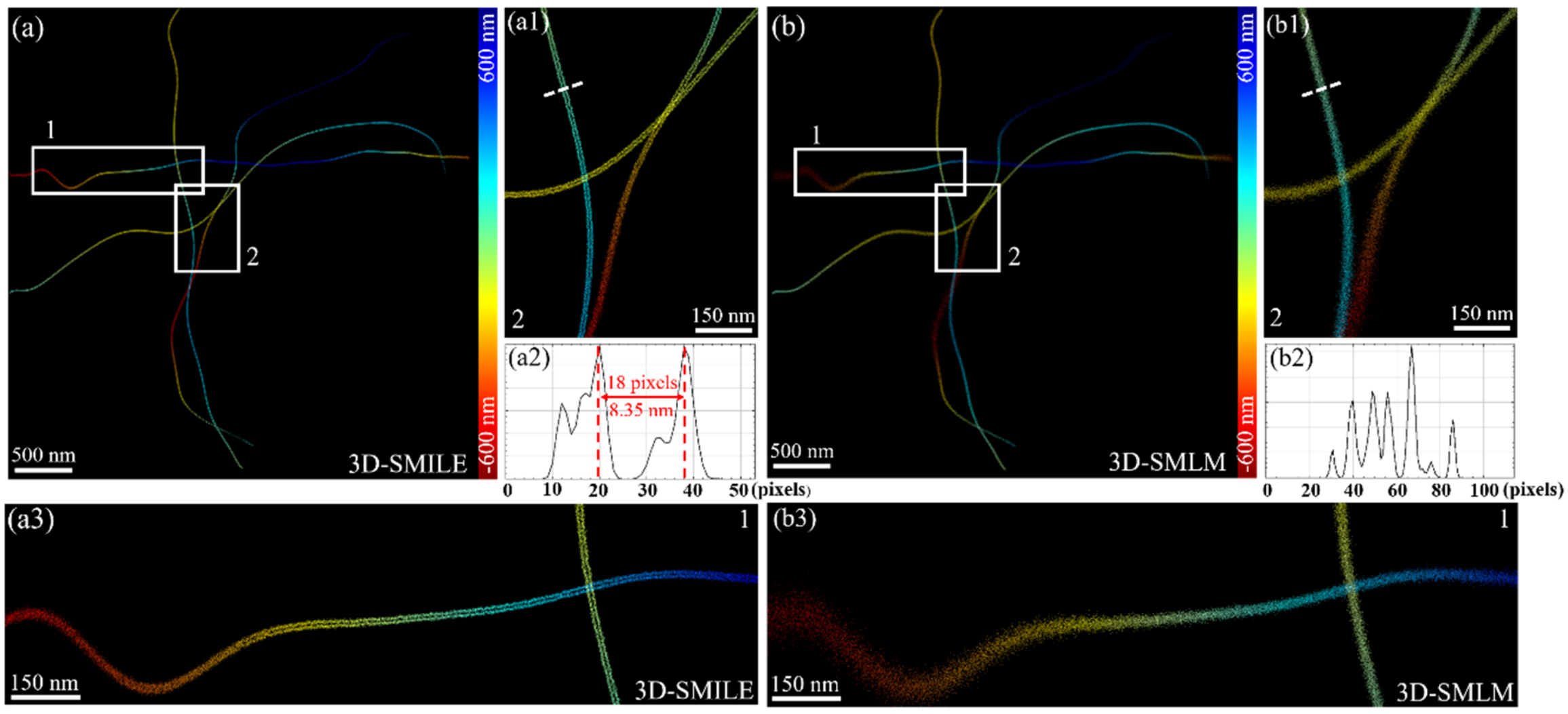

**a, b** Color-coded (depth) 3D images resulting from Ast-SMILE and 3D-SMLM, respectively. **a1, a3** Zoom-in views of the white box areas in **a**; **b1, b3** Zoom-in views of the white box areas in **b**. **a2, b2** Intensity profiles of the white dotted line in **a1** and **b1**.

In this simulation, the lateral gap between the adjacent free curves is 8.5 nm. The gap cannot be resolved in 3D-SMLM (**b1** and **b3**) but can be seen in Ast -SMILE (**a1** and **a3**), the distance between the adjacent free curves measured in Ast -SMILE is around 8.35 nm, consistent with the ground truth (8.5 nm).

**Supplementary Figure 6: Different illumination schemes for Ast SMILE**

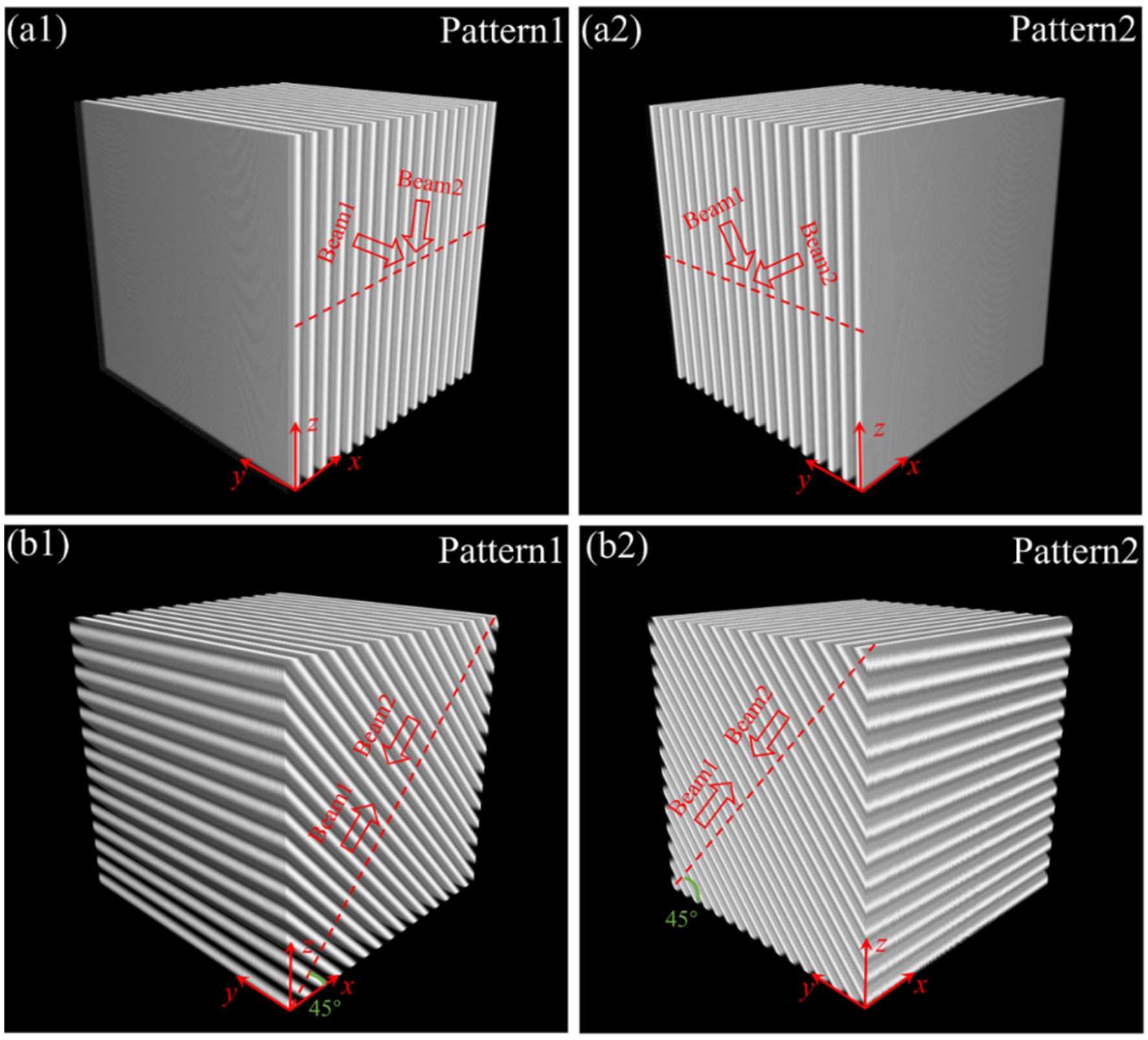

**a1, a2** Axially homogeneous illumination scheme. The red arrows indicate the incident directions of the two interference beams. To ensure the uniformity in axial direction, the incident wave vectors in z direction of the two beams need to kept close to same. **b1, b2** Oblique illumination scheme. The dihedral angle formed by the stripes and the transverse plane is 45 degrees. The red arrows indicate the incident directions of the two interference beams (better using two objective lenses for the illumination to ensure the minimum pitch size). This illumination scheme is similar to ModLoc, but ModLoc utilizes a normal incidence beam and an oblique incidence beam for interference to generate oblique patterns.

## Supplementary Note 1: Expressions of partial derivatives for the forward imaging model

Derivative of $N_x$:

$$\begin{cases} \frac{\partial I_x^i}{\partial N_x} = \frac{1}{3} \times \left(1 + m_x\, sin\left(\overrightarrow{p_x} \cdot (x_0, y_0, z_0) + (i-1) \times \frac{2\pi}{3} + \varphi_x\right)\right) \times PSF(x - x_0, y - y_0, z_0) \\ \frac{\partial I_y^j}{\partial N_x} = 0 \end{cases} \tag{S1}$$

Derivative of $N_y$:

$$\begin{cases} \frac{\partial I_x^i}{\partial N_y} = 0 \\ \frac{\partial I_y^j}{\partial N_y} = \frac{1}{3} \times \left(1 + m_y\, sin\left(\overrightarrow{p_y} \cdot (x_0, y_0, z_0) + (j-1) \times \frac{2\pi}{3} + \varphi_y\right)\right) \times PSF(x - x_0, y - y_0, z_0) \end{cases} \tag{S2}$$

Derivative of $m_x$:

$$\begin{cases} \frac{\partial I_x^i}{\partial m_x} = \frac{N_x}{3} \times sin\left(\overrightarrow{p_x} \cdot (x_0, y_0, z_0) + (i-1) \times \frac{2\pi}{3} + \varphi_x\right) \times PSF(x - x_0, y - y_0, z_0) \\ \frac{\partial I_y^j}{\partial m_x} = 0 \end{cases} \tag{S3}$$

Derivative of $m_y$:

$$\begin{cases} \frac{\partial I_x^i}{\partial m_y} = 0 \\ \frac{\partial I_y^j}{\partial m_y} = \frac{N_y}{3} \times sin\left(\overrightarrow{p_y} \cdot (x_0, y_0, z_0) + (j-1) \times \frac{2\pi}{3} + \varphi_y\right) \times PSF(x - x_0, y - y_0, z_0) \end{cases} \tag{S4}$$

Derivative of $x_0$:

$$\begin{cases} \frac{\partial I_x^i}{\partial x_0} = \frac{N_x}{3} \times \left(1 + m_x\, sin\left(\overrightarrow{p_x} \cdot (x_0, y_0, z_0) + (i-1) \times \frac{2\pi}{3} + \varphi_x\right)\right) \times \frac{\partial PSF(x-x_0, y-y_0, z_0)}{\partial x_0} \\ \qquad + \frac{N_x}{3} \times m_x \times p_{xx} \times cos\left(\overrightarrow{p_x} \cdot (x_0, y_0, z_0) + (i-1) \times \frac{2\pi}{3} + \varphi_x\right) \\ \frac{\partial I_y^j}{\partial x_0} = \frac{N_y}{3} \times \left(1 + m_y\, sin\left(\overrightarrow{p_y} \cdot (x_0, y_0, z_0) + (j-1) \times \frac{2\pi}{3} + \varphi_y\right)\right) \times \frac{\partial PSF(x-x_0, y-y_0, z_0)}{\partial x_0} \\ \qquad + \frac{N_y}{3} \times m_y \times p_{yx} \times cos\left(\overrightarrow{p_y} \cdot (x_0, y_0, z_0) + (j-1) \times \frac{2\pi}{3} + \varphi_y\right) \end{cases} \tag{S5}$$

Derivative of $y_0$:

$$\begin{cases} \frac{\partial I_x^i}{\partial y_0} = \frac{N_x}{3} \times \left(1 + m_x\, sin\left(\overrightarrow{p_x} \cdot (x_0, y_0, z_0) + (i-1) \times \frac{2\pi}{3} + \varphi_x\right)\right) \times \frac{\partial PSF(x-x_0, y-y_0, z_0)}{\partial y_0} \\ \qquad + \frac{N_x}{3} \times m_x \times p_{xy} \times cos\left(\overrightarrow{p_x} \cdot (x_0, y_0, z_0) + (i-1) \times \frac{2\pi}{3} + \varphi_x\right) \\ \frac{\partial I_y^j}{\partial y_0} = \frac{N_y}{3} \times \left(1 + m_y\, sin\left(\overrightarrow{p_y} \cdot (x_0, y_0, z_0) + (j-1) \times \frac{2\pi}{3} + \varphi_y\right)\right) \times \frac{\partial PSF(x-x_0, y-y_0, z_0)}{\partial y_0} \\ \qquad + \frac{N_y}{3} \times m_y \times p_{yy} \times cos\left(\overrightarrow{p_y} \cdot (x_0, y_0, z_0) + (j-1) \times \frac{2\pi}{3} + \varphi_y\right) \end{cases} \tag{S6}$$

Derivative of $z_0$:

$$\begin{cases} \frac{\partial I_x^i}{\partial z_0} = \frac{N_x}{3} \times \left(1 + m_x\, sin\left(\overrightarrow{p_x} \cdot (x_0, y_0, z_0) + (i-1) \times \frac{2\pi}{3} + \varphi_x\right)\right) \times \frac{\partial PSF(x-x_0, y-y_0, z_0)}{\partial z_0} \\ \qquad + \frac{N_x}{3} \times m_x \times p_{xz} \times cos\left(\overrightarrow{p_x} \cdot (x_0, y_0, z_0) + (i-1) \times \frac{2\pi}{3} + \varphi_x\right) \\ \frac{\partial I_y^j}{\partial z_0} = \frac{N_y}{3} \times \left(1 + m_y\, sin\left(\overrightarrow{p_y} \cdot (x_0, y_0, z_0) + (j-1) \times \frac{2\pi}{3} + \varphi_y\right)\right) \times \frac{\partial PSF(x-x_0, y-y_0, z_0)}{\partial z_0} \\ \qquad + \frac{N_y}{3} \times m_y \times p_{yz} \times cos\left(\overrightarrow{p_y} \cdot (x_0, y_0, z_0) + (j-1) \times \frac{2\pi}{3} + \varphi_y\right) \end{cases} \tag{S7}$$

Derivative of $bg_x^i$:

$$\begin{cases} if\ i = k,\ \frac{\partial I_x^i}{\partial bg_x^k} = 1;\ else \quad \frac{\partial I_x^i}{\partial bg_x^k} = 0 \\ \frac{\partial I_y^j}{\partial bg_x^k} = 0 \end{cases} \tag{S8}$$

Derivative of $bg_y^j$:

$$\begin{cases} \frac{\partial I_x^i}{\partial bg_y^k} = 0 \\ if\ j = k,\ \frac{\partial I_y^j}{\partial bg_y^k} = 1\ ;\ else \quad \frac{\partial I_y^j}{\partial bg_y^k} = 0 \end{cases} \tag{S9}$$

where $i, j, k = 1, 2, 3$, $\overrightarrow{p_x} = (p_{xx}, p_{xy}, p_{xz})$ and $\overrightarrow{p_y} = (p_{yx}, p_{yy}, p_{yz})$. For partial derivative of $PSF(x - x_0, y - y_0, z_0)$ with respect to $x_0/y_0/z_0$ can be solved according to the cubic-spline representation of the 3D PSF.

## Supplementary Note 2: Ratiometric calculation and estimation of 3D pattern parameters

The ratiometric calculation, i.e., calculation of photon number of each sub-image, is realized through jointly fitting of six sub-images (but without combining global information of 3D patterns). Each sub-image has same 3D coordinate but different photon number and background noise. The forward imaging model can be described as following:

$$I_i = N_i \times PSF(x - x_0, y - y_0, z_0) + bg_i \tag{S10}$$

where $i = 1\sim3$ denotes sub-images modulated by $x$ illumination and $i = 4\sim6$ denotes sub-images modulated by $y$ illumination. The parameters remaining to be determined are $[N_1\sim N_6, x_0, y_0, z_0, bg_1\sim bg_6]$, and a loss function based on MLE is constructed to solve all the parameters,

$$\chi^2_{MLE} = 2\sum_{i=1}^{6}\left(\sum_k\left(\mu_k^i - x_k^i\right) - \sum_k x_k^i ln\left(\frac{\mu_k^i}{x_k^i}\right)\right) \tag{S11}$$

where $\mu_k^i$ are the expected number of photons in pixel $k$ of $i^{th}$ sub-images which can be derived from forward imaging model combined with cubic-spline interpolated PSF model and $x_k^i$ are the measured number of photons. To minimize $\chi^2_{MLE}$, Levenberg–Marquardt (L-M) algorithm with GPU acceleration can be used, similar to what has been described in the Methods.

This fitting process provides rough position estimates $(x_0, y_0, z_0)$, and at the same time, $N_i$ indicates the relative location with respect to interference fringes. Take sub-images illuminated by patterns along x direction as an example,

$$\begin{aligned} N_i &= \frac{N_x}{3} \times \left(1 + m_x sin\left(\overrightarrow{p_x} \cdot (x_0, y_0, z_0) + \varphi_x + (i-1) \times \frac{2\pi}{3}\right)\right) \\ &= \frac{N_x}{3} \times \left(1 + m_x sin\left(\varphi + (i-1) \times \frac{2\pi}{3}\right)\right) \end{aligned} \tag{S12}$$

where $i = 1\sim3$ denotes sub-images with different phase shifts, $\overrightarrow{p_x}$, $\varphi_x$, $m_x$ and $N_x$ is the same as those in Eq. 5 in the Methods. It can be seen that $\varphi = \overrightarrow{p_x} \cdot (x_0, y_0, z_0) + \varphi_x$ is only related to the position of the molecule and pattern parameters ($\overrightarrow{p_x}$ and $\varphi_x$). $\varphi$ reflects the relative position of the molecule with respect to the intensity peak/bottom of the pattern, and ultimately affect the ratio of photon numbers between each sub-image. $N_x$, $m_x$ and $\varphi$ can be easily calculated as following,

$$N_x = N_1 + N_2 + N_3 \tag{S13}$$

$$m_x = \frac{3 \times \sqrt{\left(\frac{N_2 - \frac{N_x}{3} - \left(N_1 - \frac{N_x}{3}\right)\cos\left(\frac{2\pi}{3}\right)}{\sin\left(\frac{2\pi}{3}\right)}\right)^2 + \left(N_1 - \frac{N_x}{3}\right)^2}}{N_x} \tag{S14}$$

$$\varphi = \ arcsin\left(\frac{3N_1 - N_x}{m_x}\right) \tag{S15}$$

Here $\varphi$ ranges from $-\pi$ to $\pi$. The fringe parameters $\overrightarrow{p_x}$ and $\varphi_x$ can be estimated by combining the rough estimates $(x_0, y_0, z_0)$ as well as $\varphi$ by solving the nonlinear minimization problem. The loss function is constructed as follows:

$$\chi^2 = \ \left\| PMOD\left(\widetilde{\boldsymbol{p_x}} \cdot (x_0, y_0, z_0)\right) + \widetilde{\varphi_x} - \ \varphi \right\|^2 \tag{S16}$$

where $\widetilde{\boldsymbol{p_x}}$ and $\widetilde{\varphi_x}$ is the optimized parameters and $\varphi$ is got from Eq. S6. Function PMOD is to map the phase value to the interval from $-\pi$ to $\pi$. Built-in optimization toolbox in MATLAB is used to solve this problem ('lsqcurvefit' function using 'trust-region-reflective' algorithm).